\documentstyle[12pt]{article}

\begin{document}

\begin{titlepage}
\begin{flushright}
IJS-TP-98/25\\
NUHEP-TH-98-22\\
CPT-S695.1298\\
\end{flushright}

\begin{center}
{\Large \bf  The penguin operators in 
nonresonant $B^- \to M \bar M \pi^-$ $(M =\pi^-, K^-, K^0)$  decays }\\
\vspace{1cm}
{\large \bf S. Fajfer$^{a,b}$, R. J. Oakes$^{c}$ and  
T.N. Pham$^{d}$\\}

{\it a) J. Stefan Institute, Jamova 39, P. O. Box 3000, 1001 Ljubljana, 
Slovenia}
\vspace{.5cm}

{\it b) 
Department of Physics, University of Ljubljana, Jadranska 19, 1000 Ljubljana,
Slovenia}
\vspace{.5cm}

{\it c) Department of Physics and Astronomy, Northwestern University, 
Evanston, Il 60208, U.S.A.\\}
\vspace{.5cm}
               
{\it d) Centre de Physique Theorique, Centre National de la Recherche 
Scientifique, 
UMR 7644, Ecole Polytechnique, 91128 Palaiseau Cedex, France\\}

\end{center}

\vspace{0.25cm}

\centerline{\large \bf ABSTRACT}

\vspace{0.2cm}

We investigate the contributions coming from the 
penguin operators  
in  the nonresonant  $B^- \to  M \bar M \pi^-$ ($M =\pi^-, K^-, K^0$) decays. 
The effective Wilson coefficients of the 
the  strong penguin operators $O _{4,6}$ 
are found to be relatively larger.  
We calculate the contributions arising from the $O_4$ and $O_6$ 
operators in the  nonresonant decays $B^- \to M \bar M\pi^-$ 
($M = \pi^-, K^-, \bar K^0$) using a model combining heavy quark symmetry 
and the chiral symmetry, developed previously.   
We find that the forbidden nonresonant $B^- \to K^0 \bar K^0 \pi^-$
decay occurs 
through the strong penguin operators. 
These  penguin contributions affect the branching ratios for $B^- \to M \bar M\pi^-$
($M = \pi^-, K^-$)
by only a few percent.  The branching ratio for 
$B^- \to K^0 \bar K^0 \pi^-$  is estimated to 
be  of the order $ 10^{-6}$. 

\end{titlepage}


There is considerable 
interest in understanding the 
 decay mechanism of the
 nonleptonic charmless three body decays of B mesons 
 \cite{BBGM1,DEHT,EOS,BFOPS}. 
The importance of penguin operators in three body 
decays of charged B mesons has recently been questioned \cite{BBGM1}. 
In the analysis of the Dalitz plot 
for $B^- \to \pi^+ \pi^- \pi^-$ 
the  authors of \cite{BBGM1}
 have assumed that the nonresonant decay amplitude is flat, having no 
dependence on the Dalitz variables. They also assumed that 
the contributions of the penguin operators can amount to as much as  $20 \%$ 
of the dominant decay amplitude.  
Others have made predictions for the branching ratios of decays \cite{DEHT,EOS,BFOPS} 
motivated in part by the CLEO limits on some of the nonresonant decays 
of the type  $B ^+ \to h^+ h^+ h^-$ \cite{CLEO}. 
CLEO found the upper limits on the
branching ratios ${\rm BR}(B^+ \to \pi^+ \pi^- \pi^+) \le 4.1 \times 10^{-5}$ 
and
${\rm BR}(B^+ \to K^+ K^-\pi^+) \le 7.5 \times 10^{-5}$. 
In addition there  is hope that the CP violating phase $\gamma$ 
can be measured  from 
the asymmetry in charged B meson charmless three body decays 
\cite{DEHT,BFOPS,EGM}.

Motivated by the need to understand whether the 
nonresonant decay amplitudes for 
$B^- \to M {\bar M} \pi^-$ ($ M = \pi^+, K ^+$) involve  
significant effects due to  the penguin 
operators 
we have investigated the contributions  coming from 
the penguin operators \cite{Ali} - \cite{BDHP} in 
these nonresonant decay amplitudes. 
The  decay $B^- \to K^0 \bar K^0 \pi^-$ 
is CKM forbidden  \cite{NQ,Nir}.
However, 
we found that $B^- \to K^0 \bar K^0 \pi^-$ occurs 
through penguin operators. A measurements of this rate would  allow one  
to extract the product of the CKM matrix elements 
 $V_{tb} V_{td}^*$.

In our analysis we will  
use of the factorization approximation in which 
the main contribution to the nonresonant $B^- \to M \bar M\pi^-$ 
amplitudes come from  either the product
$< M \bar M| ({\bar u} b)_{V - A}| B^->$ $ < \pi^- | ({\bar d} u)_{V -A} |0>$ \break  
or $< \pi^- \bar M| ({\bar d} b)_{V - A}| B^->$ $ <M | ({\bar u} u)_{V -A} |0>$,  
where $(\bar q_1 q_2 )_{V-A}$
denotes  \break $\bar q_1 \gamma_{\mu} (1- \gamma_5)q_2$.
For the calculation of the matrix element
$< M \bar M| ({\bar u} b)_{V - A}| B^->$ 
we extend   the results obtained in \cite{BFOP}, where the nonresonant
$D^+ \to K^- \pi^+ l \nu$ decay
was analyzed. The experimental result for 
the branching ratio of 
the nonresonant $D^+ \to K^- \pi^+ l \nu$ decay was successfully reproduced
within a hybrid framework \cite{BFOP} which combines the
heavy quark effective theory (HQET)
and the chiral Lagrangian (CHPT) approach. 
The combination  of heavy quark symmetry and  chiral symmetry has also  been 
quite successful in other  
analyses of D meson semileptonic decays \cite{casone} - \cite{BFOS}. 

Heavy quark symmetry is expected to be even  better
for the heavier B mesons \cite{caspr,wise}. However, CHPT 
might be less reliable in B decays due to the large
energies of light mesons in 
the final state. It is really only known  that the combination of
HQET and CHPT is valid at small
recoil momentum. To take into account the larger recoil energies  
of the light mesons in our previous work \cite{BFOP,BFO}, we modified 
the hybrid model of \cite{casone} - \cite{wise}
to describe
the semileptonic decays of D mesons into one light vector or pseudoscalar 
meson. 
Our modification is quite straightforward: we  retain 
the usual HQET Feynman rules for the
{\it vertices} near and outside the zero-recoil region, 
{\it but we include the complete
propagators instead of using the usual HQET propagator}.
This quite reasonable modification  of the hybrid HQET and CHPT model enabled us to use 
it successfully  
over the entire  kinematic region of the D meson weak decays 
\cite{BFOP,BFO,BFOS}. 

In the following we systematically use this model 
to calculate the contributions of the penguin operators to the nonresonant 
$B^- \to M \bar M\pi^-$ $(M=\pi^-, K^-, K^0)$ decay amplitudes.  
We first analyze the contributions coming from the $O_{4,6}$ penguin 
operators \cite{Ali1,BDHP}, since their effective Wilson coefficients 
are the largest. We then determine the dependence on the Dalitz plot variables. 
The operator $O_4$, as defined 
 in \cite{Ali1,BDHP}, has the same dependence on the Dalitz plot variables   
as the tree-level operator  $O_1$, while  $O_6$  exhibits different energy  dependence. 
Finally, we discuss the influence of these operators on the branching ratios for  
$B^- \to M \bar M \pi^-$ $(M = \pi^+, K^+)$ and estimate the branching rate for 
$ B^- \to K^0 \bar K^0 \pi^-$.

The effective weak Hamiltonian  for the nonleptonic Cabibbo-suppressed 
$B$ meson decays is given by \cite{Ali1,Ali2,BBJ,BDHP}
\begin{eqnarray}
{\cal H}_{eff}&  = & \frac{G_F}{{\sqrt 2}} [V_{ud}^* V_{ub} (c_1 O_{1u} +
c_2 O_{2u} )  + V_{cd}^* V_{cb} (c_1 O_{1c} +c_2 O_{2c} ) ]\nonumber\\
& - &\sum_{i=3}^{10} ([V_{ub} V_{ud}^* c_i^u 
 +  V_{cb} V_{cd}^* c_i^c + V_{tb} V_{td}^* c_i^t) O_i ] + h.c.
\label{eq1}
\end{eqnarray}
where the superscripts $u$, $c$, $t$ denote the internal quark. 
The operators $O_i$ are defined in \cite{Ali1,BBJ,BDHP}. We rewrite 
$O_3$ $-$ $ O_6$, using the Fierz transformations, as follows:
\begin{equation} 
O_3 = \sum_{q = u,d,s,c,b} \bar d \gamma_{\mu} (1 - \gamma_5) b 
\bar q \gamma^{\mu} (1 - \gamma_5) q ,
\label{O3} 
\end{equation}
\begin{equation}
O_4 = \sum_{q = u,d,s,c,b} \bar d \gamma_{\mu} (1 - \gamma_5) q
\bar q \gamma^{\mu} (1 - \gamma_5) b,
\label{O4}
\end{equation}
\begin{equation}
O_5 = \sum_{q = u,d,s,c,b} \bar d \gamma_{\mu} (1 - \gamma_5) b
\bar q \gamma^{\mu} (1 +  \gamma_5) q ,
\label{O5}
\end{equation} 
\begin{equation}
O_6 = - 2 \sum_{q = u,d,s,c,b} \bar d (1 - \gamma_5) q
\bar q  (1 +\gamma_5) b,
\label{O6}
\end{equation}
We consider two possibilities for  the Wolfenstein parameters $\rho$ and $\eta$: 
Case I, the 
effective Wilson coefficients $a_i^{eff}$ 
 determined in  
\cite{Ali1} for $b \to d$ transitions  with $N_c=3$ , 
and $\rho=0.12$ and 
$\eta = 0.34$:
\begin{equation}
a_4^{eff} = -0.0412 - 0.0036 i 
\label{a_4}
\end{equation}
and 
\begin{equation}
a_6^{eff} = -0.0548 - 0.0036 i. 
\label{a_6}
\end{equation}
Case II, the effective Wilson coefficients $a_i^{eff}$ determined in  
\cite{Ali2} for  $b \to d$ transitions with  
$N_c=3$ and $\rho=0.05$ and 
$\eta = 0.45$:
\begin{equation}
a_4^{eff} = -0.048 - 0.007 i 
\label{a_42}
\end{equation}
and 
\begin{equation}
a_6^{eff} = -0.060 - 0.007 i. 
\label{a_62}
\end{equation}
The effective coefficients $a_3^{eff}$ and $a_5^{eff}$ are one order of 
magnitude smaller \cite{Ali1,Ali2} than these two 
and therefore we can safely neglect the contributions arising from 
$O_{3,5}$ operators. 

The quark currents required in the weak Hamiltonian (\ref{eq1})
can be  expressed
in terms of the meson fields, as previously described explicitly in 
\cite{BFOPS,BFOP,BFO}.
The operator $O_6$ can be rewritten as  the product of  the density operators. 
For the 
$\bar d (1 - \gamma_5) q$ scalar and pseudoscalar quark  density operator 
we use the CHPT result \cite{GL}. 
The explicit chiral symmetry breaking, to lowest order in the 
chiral expansion, is obtained by adding 
the quark mass term \cite{GL}
\begin{equation} 
{\cal L}_s = \frac{f_{\pi}^2}{4} \{ {\rm tr\,}{\cal B} (M U^{\dag} + U   
M^{\dag})\}, 
\label{CLB}
\end{equation}
where $ M = {\rm diag}( m_u, m_d, m_s)$ and ${\cal B}$ is a real constant that 
can be 
expressed in terms of quark 
and meson masses; e.g., to lowest order $m_{K^0}^2 = {\cal B} (m_s + m_d)$ and  
$U = exp (i2 \Pi /f)$ where $ \Pi$ is a pseudoscalar meson matrix. Using  
(\ref{CLB}) one can  
easily  bosonize the density operators: 
\begin{equation}
\bar q_i (1 - \gamma_5) q_j = - \frac{f_{\pi}^2}{2} {\cal B} U_{ji}
\label{den}
\end{equation}
For the calculation of the density operator $\bar q  (1 +\gamma_5) b$   we use 
the relations \cite{Ali2}
\begin{equation}
\bar q \gamma_5 b = \frac{-i}{m_b}\partial_{\alpha}(\bar q \gamma^{\alpha} 
\gamma_5 b), 
\label{hsr}
\end{equation}
and 
\begin{equation}
\bar q b = \frac{i}{m_b}\partial_{\alpha} (\bar q \gamma^{\alpha}  b), 
 \label{hsrv}
\end{equation}
where $m_q$ has been dropped since $m_q << m_b$. 

The evaluation of the matrix elements $<  M| \bar q (1 +\gamma_5) b |B>$ and 
$< M  M| \bar q (1 +\gamma_5) b |B>$ can then be reduced to 
the evaluation of the matrix elements 
of the weak currents  
$< MM  | \bar q \gamma_{\mu} \gamma_5 b |B>$ and  $< M | \bar q \gamma_{\mu}  b |B>$.  
 Assuming factorization, we  evaluate the matrix elements of the  
operator $O_6$:
\begin{eqnarray}
<M M M| O_6| B> & = & -2  \sum_{u,d,s,c,b} \{
< M |\bar d (1- \gamma_5) q |0 > <M M | \bar q (1 + \gamma_5) b |B> \nonumber\\
& + & < M M|\bar d (1- \gamma_5) q |0 > < M | \bar q (1 + \gamma_5) b 
|B>\nonumber\\
& + & < MM M|\bar d (1- \gamma_5) q |0 > < 0 | \bar q (1 + \gamma_5) b |B>\}.
\label{o6fac}
\end{eqnarray}

The matrix elements $< M |\bar d (1- \gamma_5) q |0 >$, 
$< MM|\bar d (1- \gamma_5) q |0 >$, and \break 
 $<MMM|\bar d (1- \gamma_5) q |0 >$ 
are easily calculated using (\ref{den}). 
For the calculation of the matrix elements $<M | \bar q \gamma_{\mu}b |B>$ 
and  $<M M | \bar q \gamma_{\mu} \gamma_5 b |B>$  
we generalize  the results obtained in the  
analysis of $D$ meson semileptonic decays described in detail in \cite{BFOP}  
and \cite{BFO}.
The matrix element $< M| \bar q \gamma_{\mu} (1- \gamma_5)b |B>$ is 
given by
\cite{BFO,BFOS}
\begin{eqnarray}
< M(p')| \bar q \gamma_{\mu} (1- \gamma_5)b|B(p_B)> & = & [ (p_B + p')_{\mu}  - 
\frac{m_H^2 - m_M^2}{q^2} q_{\mu}] F_1(q^2) \nonumber\\
& + & \frac{m_H^2 - m_M^2}{q^2} q_{\mu} F_0(q^2),
\label{ffbp}
\end{eqnarray} 
where $q = p_B - p'$ and $F_1(0) = F_0(0)$. 
The form factors are found to be \cite{BFO,BFOS}
\begin{eqnarray}
F_1(q^2) & = & - \frac{f_B}{2} + g  f_{B'*} 
\frac{m_{B'*}^{3/2} m_B^{1/2}}{q^2 - m_{B'*}^2}, 
\label{ff1}
\end{eqnarray}
and 
\begin{eqnarray}
F_0(q^2) & = & - \frac{f_B}{2} - g  f_{B'*} {\sqrt \frac{m_B}{m_{B'*}}}
\nonumber\\
& +& \frac{q^2}{m_B^2 - m_M^2} [- \frac{f_B}{2} + 
 g  f_{B'*} 
{\sqrt \frac{m_{B'*}}{ m_B}}], 
\label{ff0}
\end{eqnarray}
where $B'^*$ denotes the relevant  vector meson pole  and $g$ is  the $B^* B M$ 
coupling constant.  

To evaluate  the 
matrix element $< M_1 (p_1) M_2 (p_2) | ({\bar q_i} b)_{V -A} 
| B^- (p_B)>$
we will also use and generalize the results obtained previously in the analysis of the nonresonant 
$D^+ \to \pi^+ K ^- l \nu_l$ decay width \cite{BFOP}. 
We write the matrix element 
$< M_1 (p_1) M_2 (p_2) | ({\bar u} b)_{V-A} | B^- (p_B)>$ 
in the general form 
\begin{eqnarray}
\label{wwh}
< M_1 (p_1) M_2 (p_2) | {\bar u} \gamma_{\mu} (1 - \gamma_5) b | B^- (p_B)>
&\!\!\! = \!\!\!&
ir(p_B-p_2-p_1)_\mu\nonumber\\
+iw_+(p_2+p_1)_\mu+iw_-(p_2-p_1)_\mu
&\!\!\! -\!\!\!&2h\epsilon_{\mu\alpha\beta\gamma}p_B^\alpha p_2^\beta 
p_1^\gamma\;.
\end{eqnarray}
The form factors $w_\pm^{nr}$  for the nonresonant decay are given in 
\cite{BFOPS,BFOP}: 
\begin{eqnarray}
w_+^{nr}(p_1,p_2) & = & - \frac{g}{f_1 f_2}
\frac{f_{B*} m_{B*}^{3/2} m_B^{1/2}}{(p_B - p_1)^2 - m_{B*}^2} [ 1 - 
\frac{p_1\cdot (p_B - p_1)}{2 m_{B*}^2} ]\nonumber\\
& + & \frac{f_B}{ 2 f_1f_2}  - \frac{{\sqrt m_B} \alpha_2}{ 2 f_1 f_2}
\frac{1}{m_B^2}p_B\cdot (p_2 - p_1 ),
\label{w+1}
\end{eqnarray}
\begin{eqnarray}
w_-^{nr}(p_1,p_2) & = &  \frac{g}{f_1f_2}
\frac{f_{B*} m_{B*}^{3/2} m_B^{1/2}}{(p_B - p_1)^2  - m_{B*}^2}
[ 1 + \frac{p_1\cdot (p_B - p_1)}{2 m_{B*}^2}]\nonumber\\
&  + &\frac{{\sqrt m_B} \alpha_1}{f_1f_2}.
\label{w-1}
\end{eqnarray}
The parameters $\alpha_{1,2}$ are 
 defined in \cite{BFO}. Note that both the $\alpha_1$ and $\alpha_2$
terms are important in (\ref{w+1}) and (\ref{w-1}),
which was previously overlooked  \cite{DEHT}.
Within this same framework \cite{BFOP,BFO} we evaluate $r^{nr}$
\begin{eqnarray}
r^{nr}(p_1, p_2) & = & - \frac{ 1 + \tilde \beta}{f_1 f_2} 
p_B\cdot (p_2 - p_1) 
{\sqrt \frac{m_{B'}}{m_B}} \frac{f_{B'}}{(p_B - p_1 - p_2)^2 - m_{B'}^2} 
\nonumber\\
& - & {\sqrt \frac{m_B}{m_{B}}} 
\frac{ 4 g^2 f_{B''} m_{B'}^* m_{B'}}{f_1 f_2} 
\frac{1}{(p_B - p_1 - p_2)^2 - m_{B''}^2}
 \nonumber\\
& \times & \frac{[p_1 \cdot p_2 - \frac{1}{m_{B*'}^2} p_2 \cdot (p_B - p_1) 
p_1\cdot (p_B - p_1)]}{(p_B - p_1)^2 - m_{B'*}^2} \nonumber\\
& + & \frac{2 g}{f_1 f_2} \frac{f_{B'^*} m_{B'^*}}{(p_B - p_1 )^2-  m_{B'*}^2} 
\frac{p_1\cdot (p_B - p_1)}{ m_{B*'}^2} \nonumber\\
& + & \frac{f_B}{2f_1 f_2} + \frac{\alpha_2  {\sqrt m_B}}{2 m_B^2} 
p_B \cdot (p_2 - p_1)
\label{r}
\end{eqnarray}
Here  $ B'$, $B^{'*}$, $B''$ denote the relevant  $B$ meson poles, and $f_{1,2}$ 
 denotes the pseudoscalar meson decay constants. The coupling 
$\tilde \beta $ has been analyzed in \cite{BFOS} and found to be close to zero  
and therefore will be neglected. 

The matrix element of the operator $O_4$ can be evaluated straightforwardly 
using factorization:
\begin{eqnarray} 
< \pi^- \pi^+ \pi^-| O_4| B^- >_{nr}& =& 
<\pi^+ \pi^-| \bar u\gamma_{\mu}(1- \gamma_5) b | B^->_{nr}\nonumber\\
&\enspace& < \pi^- | \bar d \gamma^{\mu}(1- \gamma_5)
 u |0>,
\label{o4-1}
\end{eqnarray}
and the corresponding expression for the 
$B^- \to   K^- K^+ \pi^-$ matrix elements can simply be obtained  
by the replacement $\pi^+$ and $ \pi^- $ by $K^+$ and $K^-$ respectively. 
Note that the matrix element $< MM| \bar q_1 \gamma_{\mu}(1- \gamma_5) q_2  |0>$ 
 is dominated by resonant contributions. 
Using the variables  $s =(p_B - p_3)^2 = (p_2+ p_1)^2$, $t =(p_B- 
p_1)^2 = (p_2 + p_3)^2$
$u = (p_B - p_2)^2 = (p_1+ p_3)^2$ and 
the  pseudoscalar meson decay constants  $f_i$, we can then write the nonresonant decay 
matrix element of $O_4$ as
\begin{eqnarray}
< M_1(p_1) M_2 (p_2) \pi^- (p_3)| O_4| B^- (p_B)>_{nr} & = & [ f_\pi m_{\pi}^2 
r^{nr} (s,t) + 
 {f_{\pi} \over 2} (m_B^2 \nonumber \\
 -s -m_{\pi}^2) w_+^{nr} (s,t)  +  
 {f_{\pi} \over 2} (2 t & + &s - m_B^2 - m_1^2 - m_2^2  
-   m_{\pi}^2) w_-^{nr}(t) ] \nonumber \\
\label{O4me}
\end{eqnarray}
where $M_1$, $M_2$ represent either $\pi^+$, $\pi^-$  or $K^+$, $K^-$.

Using  factorization the matrix elements of $O_6$ can be written as
\begin{eqnarray}
< \pi^+(p_1) \pi^-(p_2) \pi^-(p_3)|O_6  |B^-(p_B) >_{nr}&  = & \nonumber\\ 
- 2 
 \{ < \pi^+(p_1) \pi^-(p_2) | \bar u \gamma_{\mu} b| B^-(p_B)> & \times&
 \nonumber\\
 \frac{p_3^{\mu}}{m_B} 
< \pi^-(p_3)| \bar d (1 - \gamma_5) u |0> & + &\nonumber\\ 
< \pi^+(p_1) \pi^-(p_2) \pi^-(p_3)| \bar d (1 - \gamma_5) u |0> & 
\times&\nonumber\\
<0| \bar u \gamma_{\mu} b| B^-(p_B)> \frac{p_B^{\mu}}{m_B}& +&\nonumber\\
 < \pi^+(p_1) \pi^-(p_2)| \bar d (1 - \gamma_5) u |0> 
<\pi^-(p_3)| \bar u \gamma_{\mu} b| B^-(p_B)>&\times&\nonumber\\
\frac{p_B^{\mu}  -p_{3}^{\mu}}{m_B}
 + (p_1 \leftrightarrow p_3)\},  
\label{06-0}
\end{eqnarray}
where we have assumed  $m_b \simeq m_B$. 
The corresponding result for  $B^- \to \pi^- K^+ K^-$ can be straightforwardly 
obtained simply by replacing $\pi^+ \pi^- $ by $K^+ K^-$. 
Using the expressions for the matrix elements of the current and  
the density operators  we find 
\begin{eqnarray}
<  M_1(p_1) M_2(p_2) \pi^- (p_3) |O_6| B (p_B)>_{nr} &  = & 
-  f_{3}\frac{ {\cal B}}{m_B} \nonumber\\
\{\frac{1}{2}[m_3^2  r^{nr} + ( m_B^2   -  m_3^2 - s) + 
w_+^{nr}( 2 t + s - m_B^2 - m_1^2 - m_2^2 & - & m_3^2) w_-^{nr}]  \nonumber\\
+ [( m_B^2 - m_3^2 ) F_0(s)]
- \frac{4}{3} \frac{f_3 f_B}{f_1 f_2} m_B \}. &\enspace&
\label{o6}
\end{eqnarray}
For the $B^- \to \pi^- \pi^+ \pi^-$ decay there is an additional term with 
the replacement $s \leftrightarrow t$, since there are two identical pions in the 
final state in this case. 

The nonresonant amplitudes for  the $B^- \to M \bar M \pi^-$ $(M = \pi^-, K^-)$ 
 decays can be written in terms of the following matrix elements 
\begin{eqnarray}
{\cal M}_{nr} (B^- \to M \bar M \pi^-) & = & 
\frac{G_F}{{\sqrt 2}}\{  V_{ub} V_{ud}^* 
a_1^{eff}< M \bar M \pi^- |O_1| B^-> \nonumber\\
+ V_{tb} V_{td}^*
(a_4^{eff} < M \bar M  \pi^- |O_4| B^-> & + & 
a_6^{eff} <M \bar M  \pi^- |O_6| B^->) \}.
\label{ammmp}
\end{eqnarray}
The matrix element $< M \bar M \pi^- |O_1| B^-> $ $(M = \pi^-, K^-)$ was  
analyzed in \cite{BFOPS}. 

Contrary to the CKM-allowed cases in which 
the main contribution comes  from 
the  operator $O_1$, we notice that the CKM-forbidden decay 
 $B^- \to K^0 \bar K^0 \pi^-$ 
occurs through  the  contributions of  
 the penguin operators $O_4$ and $O_6$. The nonresonant 
 matrix elements are  
\begin{eqnarray} 
<K^0 (p_1) \pi^- (p_1)  \bar K^0 (p_2) \pi^- (p_3)  |O_4| B^- (p_B) >_{nr} & = & 
\frac{f_K}{2} [  m_K^2 r^{nr} (s,t) + (m_B^2\nonumber\\
 -t - m_K^2) w_+^{nr} (s,t) + 
(2 t + s - m_B^2 - 2 m_K^2 & - & m_{\pi}^2) w_-^{nr} (s,t)]
\label{o4kko}
\end{eqnarray}
and 
\begin{eqnarray}
< K^0 (p_1) \bar K^0 (p_2) \pi^- (p_3) |O_6| B^- (p_B) >_{nr} & = &
-  {\cal B} \frac{f_K}{m_B} \{ r^{nr}(s,t) m_K^2 \nonumber\\
w_+^{nr} (s,t) (m_B^2  -  t - m_K^2)  + 
w_-^{nr} (s,t)(2 t + s &-& m_B^2 - 2 m_K^2  -  m_{\pi}^2) \nonumber\\
+ [ (m_B^2 - m_\pi^2)  F_0(s)]
&-&\frac{4}{3} \frac{f_\pi f_B}{f_K^2} m_B \}
\label{o6kko}
\end{eqnarray}
The nonresonant amplitude for the  
$B^- (p_B)\to K^0(p_1) \bar K^0 (p_2) \pi^-$ decay 
can be written in terms of these matrix elements (\ref{o4kko}) $-$ (\ref{o6kko}): 
\begin{eqnarray}
{\cal M}_{nr} (B^- \to K^0 \bar K^0 \pi^-) & = & 
\frac{G_F}{{\sqrt 2}} V_{tb} V_{td}^* \nonumber\\
\{a_4^{eff} < K^0 \bar K^0 \pi^- |O_4| B^-> & + & 
a_6^{eff} < K^0 \bar K^0 \pi^- |O_6| B^->\}.
\label{amoop}
\end{eqnarray}

The partial widths  for the nonresonant decay
$B^- \to M \bar M \pi^- $ ($M= \pi^-, K^-, K^0$) is  given by 
\begin{equation}
\Gamma_{nr} (B^- \to M \bar M \pi^-) = \frac{1}{(2 \pi)^3}
\frac{1}{32 m_B^3} \int |{\cal M}_{nr}|^2~  ds~ dt.
\label{dw}
\end{equation}

In the numerical calculation of the branching 
ratios we follow the discussion of the input parameters given in \cite{BFOPS}. 
From heavy quark symmetry we have used ${f_B}/{f_D} = $ 
$ \sqrt{{m_D}/{m_B}}$  with  the  reasonable 
choice $f_D \simeq 200$ $ \;\rm MeV$ \cite{BFOS,GM}. 
The $B$ decay  constant is then $f_B \simeq 128$ $ \;\rm MeV$.
In \cite{BFOPS}  we found that the parameters 
$\alpha_1^{B\rho } = -0.13$ $\;\rm GeV ^{1/2}$ and $\alpha_2^{B\rho} = -0.36$ 
$\;\rm GeV ^{1/2}$ 
lead to the 
branching ratio ${\rm BR}(B^- \to \pi^- \pi^+ \pi^+)$ being smaller than the 
experimental upper limit \cite{CLEO} and we rejected this possibility. 
Here we also 
use the values 
of $\alpha_{1,2}$ as in \cite{BFOPS}.  And, as discussed in \cite{BFOPS}, here  
we also  consider the range $0.2 \leq g \leq 0.23$.

In our numerical calculations we considered  both cases I and II: the 
effective coefficients $a_{4,6}^{eff}$ given in (\ref{a_4}), (\ref{a_6}) from  
\cite{Ali1} (case I), 
 and $a_{4,6}^{eff}$ given in 
 (\ref{a_42}), (\ref{a_62})  from  \cite{Ali2} (case II).  
Then for the CKM matrix we must use  the corresponding input parameters  
in the Wolfenstein parametrization of the CKM matrix 
( $V_{ub} = A \lambda^3 (\rho - i 
\eta)$, $V_{ud} = 1- 
\lambda^2/2$, $V_{td} =  A \lambda^3 (1 - \rho - i \eta)$, $V_{tb} = 1$).  
The numerical value of ${\cal B}$ can be determined from 
${\cal B}=  (2 m_K^2 - m_{\pi}^2)/2 m_s$.
Taking $m_s (\mu = 5 \;\rm GeV )  = 150$ $\;\rm MeV$, 
the same value used in \cite{Ali1} for the  
extraction of  the effective Wilson coefficients,  we find 
 ${\cal B} =1.6$ $ \;\rm GeV $. 
Inspection of the contributions coming from the $O_{4,6}$ operators 
shows some cancellations occur among the combinations  
$a_4^{eff} O_4$ 
and $a_6^{eff} O_6$. One also can explicitly see the dependence 
on the Dalitz variables. 

In Table I we present the 
penguin contributions of the operators $O_{4,6}$ to branching ratios for the 
$B^- \to \pi^- \pi^+ \pi^-$ and $B^- \to K^- K^+  \pi^-$  together with the dominant 
tree level contribution of the operator $O_1$. Both numerical results I and II 
as well as the range of g, as discussed above are presented. 

It is clearly evident from these quantitative numerical  results that the 
uncertainties 
coming from the input parameters give much larger 
uncertainties in 
the branching ratios  than the contributions of the penguin 
operators. 
Interestingly the penguin contributions, while small, are less sensitive to the input 
parameters than the dominant tree level contributions, which is quite sensitive to 
these input  parameters.
Since the amplitudes for the $B^- \to \pi^+ \pi^- \pi^+$ and $B^- \to K^+ K^- 
\pi^+$ decays 
receive rather small corrections from the penguin operators 
we do not expect significant changes in the 
CP violating asymmetry, which we 
have discussed in \cite{BFOPS}. 

We also calculated the branching ratio for the forbidden nonresonant 
decay $B^- \to K^0 \bar K^0 \pi^-$ finding 
\begin{equation}
8.4 \times 10^{-7} \le {\rm BR}(B^- \to K^0 \bar K^0 \pi^-) \le 8.7 \times 
10^{-7},
\label{brfor}
\end{equation}
for the Case I (see (\ref{a_4}) and (\ref{a_6})) and  the range $0.2 \le g \le 
0.23$.
For the Case II (see (8) and  (9))  
we found 
\begin{equation}
1.4 \times 10^{-6} \le {\rm BR}(B^- \to K^0 \bar K^0 \pi^-) \le 1.5 \times 
10^{-6}, 
\label{brfor2}
\end{equation}
for the range $0.2 \le g \le 0.23$. 
Measurement of this branching ratio is important as it provides information about the 
effective Wilson coefficients $a_{4,6}^{eff}$. 
It is interesting to note 
that in the factorization approximation, as mentioned earlier, this decay is entirely 
induced by the penguin interactions. Final state interactions (FSI)  effects, 
could alter this, however we  
believe this is unlikely  as data on 
color-suppressed decays indicate that the branching ratio for 
$B^0$ to $D^0$ and a neutral light hadron is indeed suppressed. 
Thus we expect FSI would contribute at most a branching ratio for 
$K^0 \bar K^0$ mode of the same order as the penguin terms. 
This could be checked in future measurements of this decay rate. 

 \vspace{0.5cm}
 
To summarize, we have quantitatively analyzed the penguin  contributions to 
the nonresonant 
$B^- \to M \bar M \pi^-$ decay amplitudes ($M = \pi^-, K^-, K^0$), including the 
dependence on the Dalitz variables.  
We calculated the branching ratios for 
$B^- \to M \bar M \pi^-$ decays ($M = \pi^-, K^-$) 
including the penguin contributions and 
found that they can possibly 
change the branching ratio as much as $15\%$. However, while the penguin 
contributions are small and not very sensitive to the uncertainties 
in the input parameters, the corresponding uncertainties in the dominant tree level 
contributions are considerably larger than the penguin contributions. 
We also found that the branching ratio for the CKM forbidden nonresonant 
decay $B^- \to K^0 \bar K^0 \pi^-$ is of the order $10^{-6}$ and is entirely 
induced by penguin effects. 

\vspace{1cm}

This work was supported in part by the Ministry of Science and Technology 
of the Republic of Slovenia (S.F), and by the U.S. 
Department of Energy, Division of High Energy Physics under grant 
No. DE-FG02-91-ER4086 (R.J.O.).                    
S.F. thanks the Department of Physics and Astronomy 
at Northwestern University for warm hospitality during her stay there where 
part of this work has been done.  
\newpage 
\begin{table}[h]
\begin{center}
\begin{tabular}{|c|c|c||c||c|}
\hline
$\rho$ & $\eta$ & $g$ & ${\rm BR}(B^- \to \pi^- \pi^+ \pi^-)$ & 
$ {\rm BR}(B^- \to K^+ K^- \pi^-) $ \\
\hline \hline
$0.12$ & $ 0.34$& $0.2$ & $(3.1 - 0.4) \times 10^{-5}$ & 
$(5.3-0.1) \times 10^{-5}$  \\
\hline  
$0.12$ & $ 0.34$& $0.23$ & $(3.5 - 0.5) \times 10^{-5}$ & 
$(5.8.-0.1) \times 10^{-5}$ \\
\hline\hline
$0.05$ & $ 0.45$& $0.2$ & $(4.9-0.4) \times 10^{-5}$ & 
$(8.7- 0.1) \times 10^{-5}$  \\
\hline  
$0.05$ & $ 0.45$& $0.23$ & $(5.5-0.4) \times 10^{-5}$ & 
$(9.0-0.03) \times 10^{-5}$  \\
\hline  
\end{tabular}
\caption{\label{tab1} The branching ratios for $B^- \to M \bar M \pi^-$, 
$(M = \pi^-, K^-) $ for two cases of $\rho$ and $\eta$ parameters 
determined in [8] and [9], described in the text, respectively. 
 The first number in the 
 brackets is the main contribution 
 coming from the 
 operator $O_1$ and the second number is the contribution to  
 the branching ratio 
 coming from  the operators $O_{4,6}$. 
 } 
 \end{center}
\end{table}

\end{document}